\documentclass[twocolumn,showpacs,preprintnumbers,amsmath,amssymb]{revtex4}

\usepackage{graphicx}
\usepackage{dcolumn}
\usepackage{bm}

\begin{document}
\title{Comment on ``Theory of Spin Hall Effect: Extension of the Drude Model''}
\author{V.~Ya. Kravchenko}
\affiliation{Institute of Solid State Physics, Russian Academy of Sciences, Chernogolovka, 142432 Moscow District,
Russia}
\begin{abstract}
E.~M. Chudnovsky has recently posted on the ArXive his Reply \cite{arxivshu} to my Comment \cite{0} on his recent Letter \cite{1}.  In order to avoid possible misunderstandings in the evaluation of his Letter, I am posting my Comment with some remarks on the contents of his Reply.
\end{abstract}
\maketitle

The Letter ``Theory of Spin Hall Effect: Extension of the Drude model'' \cite{1} is related to a subject that has been widely discussed during the last few years (for a review, see Ref.\cite{2}).  A simple physical model of the spin Hall (SH) effect is proposed that allegedly allows evaluating its magnitude.  In order to determine the SH component of the current driven in a metal or semiconductor by an electric field, the Author includes in the single-electronic Hamiltonian the relativistic expression of spin-orbit interaction in its conventional form, which contains $\nabla U$, the gradient of the electron's potential energy $U$. According to Eq.(2) and the explanatory text in Ref.\cite{1}, the Author ignores in $U$, without any justification, the Hartree potential created by free electrons system maintaining electrical neutrality. This omission is crucial for the final result.  After some manipulations with $\hat{p}$- and $\hat{r}$-operators, the Author proceeds with the Drude-like semiclassical equations for electron motion, \textit{i.e}., the Drude equation (11), where $-\nabla U$ is playing the role of a nonrelativistic force and $\partial^2U/\partial r_i\partial r_j$ --- the role of spin-orbit force (see Eqs.~(8), (11), and (14) in Ref.\cite{1}).  Later in the text, a very simplified situation is considered, with all conventional mechanisms of SH effect (``intrinsic'', related to the symmetry properties of the spin-orbit part of the dispersion law, and ``extrinsic'', caused by the asymmetry of impurity scattering \cite{2}) disregarded. Therefore, the SH effect proposed in Ref.\cite{1} is supposed to be of a new type.

However, the results of the Author's calculations are incorrect.

According to Eqs.~(15), (22), and (25) in Ref.\cite{1}, the SH conductivity  $\sigma_s$ is proportional to  $<\partial^2\Phi_0/\partial r_i\partial r_j>=\delta_{ij}\nabla^2\Phi_0/3$.  Here $\Phi_0$ is the \textit{total} electrostatic potential in a perfect crystal (see Eq.~(2) in Ref.\cite{1}). Therefore, it must be equal to the sum of the potentials created by the lattice and the Hartree potential of free electrons.  According to the Poisson equation, $\nabla^2\Phi_0$ is proportional to the \textit{total} electric charge density. For this charge density, the Author uses \textit{the density of the ion charge only}.  Clearly, this is incorrect. In conductors, electrons move in the self-consistent field of ions and the electron reservoir. The standard method of accounting for electron-electron interaction is insuring the electrical neutrality. In conductors, in the framework of a semiclassical approach, the condition of the neutrality is fulfilled.  Therefore, the total charge density in Eqs.~(22) and (26) is equal to zero, and SH conductivity $\sigma_s$ (see Eq.(22), (28) in Ref.[1]) vanishes.

In the Author's Reply \cite{arxivshu}, there is no justification of the complete violation of the electroneutrality condition in the considered physical situation, in the framework of the single-electron semiclassical approach. The Author actually does not respond to my criticism regarding his zero-result and simply repeats the content of his paper \cite{1}.

In conclusion, traditional mechanisms of SH effect rely on the strong microscopic electric fields in a close vicinity of nuclei. The ``new'' effect is ascribed to an ultrastrong smooth electric field developing at the sample scale. However, because of the electric neutrality, this field vanishes. The procedure of Ref.\cite{1} results in zero spin Hall effect, contrary to the Authors assertion.

\end{document}